\documentstyle[12pt]{article}

\begin{document}
\pagenumbering{arabic}
\normalbaselines
\vskip -10pt
\baselineskip=.175in

\noindent
{\bf Anderson} {\it et al.} {\bf reply (to the comment by Katz on 
``Indication, from Pioneer 10/11, Galileo, and Ulysses Data, 
of an Apparent Anomalous, Weak, Long-Range Acceleration").}

\baselineskip=.175in
\begin{quotation}
We conclude that Katz's proposal (anisotropic heat reflection off of the 
back of the spacecraft high-gain antennae, the heat coming from the RTGs)
does not provide enough power and so can not explain the Pioneer anomaly.
\end{quotation}
\baselineskip=.3in

In his comment \cite{katz}, Katz  proposes that the anomalous 
acceleration \cite{uncert} seen in the Pioneer 10/11 spacecraft 
\cite{anderson} is due to anisotropic heat reflection off of the back  
of the spacecraft high-gain antennae, the heat coming from the RTGs.  

Before launch the four RTGs delivered a total electrical power 
of 160 W (now $\sim$ 70-80 W), 
from  a total thermal fuel inventory of 2580 W (now $\sim$ 2090 W).   
Presently  $\sim 2000$ W of RTG heat must be dissipated.  
Only $\sim 75$ W  of directed power could explain the 
anomaly \cite{s}. Therefore, in principle there is enough 
power to explain the anomaly this way.  However, 
1) the geometry of the spacecraft and 2) the radiation pattern preclude it.

Many years ago this problem was discussed  with John W. Dyer, who was a 
Pioneer Project engineer at NASA/ARC, and with James A. Van Allen.  
What comes below is at least a partial reconstruction
of those discussions, which we wish to acknowledge.  

1) SPACECRAFT GEOMETRY.  The RTGs are located at the end of booms, and 
rotate about the craft in a plane that contains the approximate base of 
the antenna.  From the RTGs the antenna is thus seen ``edge on" and  
subtends a solid angle of $\sim$ 1.5 \% 
of $4\pi$ steradians \cite{ss}.   This already means  
the proposal could provide at most $\sim 30$ W.  But there 
is more. 

2) RADIATION PATTERN.  
The above estimate is based on the assumption that  
the RTGs are spherical black bodies.  But they are not. 
The main bodies of the RTGs are cylinders and they 
are grouped in two packages of two.  Each 
package has the two cylinders end to end extending away from 
the antenna. 
Every RTG has six fins that go radially out from the cylinder  
Thus, the fins are ``edge on" to the antenna (the fins point 
perpendicular to the cylinder axes).  Ignoring edge 
effects, this means that only 2.5 \% of the surface area of the RTGs  
is facing the antenna.  Further, for better radiation from the fins,  
the Pioneer SNAP 19 RTGs had larger fins than the earlier test models, and 
the packages were insulated so that the end caps had lower temperatures 
and radiated less than the cylinder/fins \cite{tele}.
As a result, the vast majority of the RTG heat is symmetrically 
radiated to space, unobscured by the antenna.

We conclude that Katz's proposal  does  not provide enough power 
and so can not explain the Pioneer anomaly \cite{addmurphy}.    

Independent of the above, we continue to search for a systematic 
origin of the effect.    

A few weeks after 
our letter \cite{anderson} was accepted, we began using new JPL 
software (SIGMA) to reduce the Pioneer 10 Doppler data to 50-day 
averages of acceleration, extending from January 1987 to July 1998, 
over a distance interval from 40 to 69 AU. 

Before mid-1990, the spacecraft rotation rate changed (slowed) by  
about -0.065 rev/day/day. Between mid-1990 and mid-1992 the 
spin-deceleration increased to -0.4 rev/day/day. But after mid-1992 the 
spin rate remained $\sim$ constant.   
In units of 10$^{-8}$ cm/s$^{2}$, the mean acceleration levels obtained 
by SIGMA from the Doppler data in these periods are \cite{uncert}:
$(7.94 \pm 0.11)$ before mid-1990, $(8.39 \pm 0.14)$ between mid-1990 
and mid-1992, and $(7.29 \pm 0.17)$ after mid-1992. 
[Similar values $(8.27\pm 0.05, ~8.77\pm 0.04, ~7.76\pm 0.08)$ 
were obtained using CHASMP.] 
We detect no long-term deceleration changes from mid-1992 to mid-1998, 
and only two spin-related discontinuities over the entire data period.

Assume that the slowing of the spin rate was caused by spacecraft 
systems (perhaps gas leak changes) that also account for a few \%
systematic effect. 
Then, excluding other biases  
(such as the radio beam decreasing the measured anomaly),
we should adopt the post-1992 value as the most accurate measure of 
the anomalous Pioneer 10 acceleration.

This work was supported by the Pioneer Project, NASA/Ames Research Center,
and was performed at the Jet Propulsion Laboratory, California Institute 
of Technology, under contract with 
NASA. P.A.L. and A.S.L. acknowledge support by a grant from NASA
through the Ultraviolet, Visible, and Gravitational Astrophysics Program.
M.M.N. acknowledges support by the U.S. DOE and the Alexander von 
Humboldt Foundation.

\vskip 20pt
\baselineskip=.175in

\noindent John D. Anderson,$^a$ Philip A. Laing,$^b$ Eunice L. Lau,$^a$ 
Anthony S. Liu,$^c$ Michael Martin Nieto,$^{d,e}$ and Slava G. Turyshev$^a$

\vskip 10pt
\baselineskip=.175in

\noindent {$^{a}$Jet Propulsion Laboratory, California Institute of 
Technology, Pasadena, CA 91109}    \\
\noindent {$^b$ The Aerospace Corporation, 2350 E. El Segundo Blvd., 
El Segundo, CA 90245-4691} \\
\noindent {$^c$ Astrodynamic Sciences, 2393 Silver Ridge Ave., Los Angeles, 
CA 90039} \\
\noindent {$^d$ Theoretical Division (MS-B285), Los Alamos National 
Laboratory, University of California, Los Alamos, NM 87545}  \\
\noindent {$^e$Abteilung f\"ur Quantenphysik, Universit\"at Ulm, 
D-86069 Ulm, Germany}

\vskip 10pt
\noindent Received \today \\
PACS numbers:  04.80.-y, 95.10.Eg, 95.55.Pe
\vskip 10pt
\baselineskip=.33in



\begin{thebibliography}{44}
\bibitem{uncert}  In understanding the significance of the 
``constancy" \cite{katz} of the data, one must 
be careful to distinguish between $\sigma(x)$ and $\langle x \rangle$.
For a Gaussian with width $a$ and centroid $x_o$, 
$\sigma(x) = a/\sqrt{2}$ can be large even though the accuracy of 
the determination of $x_o$ can be quite good since it goes as $N^{-1/2}$, 
$N$ the number of samples.
\bibitem{addmurphy}  After the early eprint \cite{katz}, Katz 
added Murphy's dissipated electrical power mechanism to his.  
We deal with that and Ulysses in our reply to Murphy.
\end{thebibliography}
\end{document}